\documentclass[preprint,aps,preprintnumbers,amsmath,amssymb,superscriptaddress]{revtex4}
\usepackage[a4paper, total={6.5in, 9in}]{geometry}
\usepackage[colorlinks=true,linkcolor=black, citecolor=blue, urlcolor=blue, 
    unicode=true]{hyperref}

\usepackage{amsmath}
\usepackage{blindtext}
\usepackage[version=4]{mhchem}
\usepackage{graphicx}
\usepackage{ulem}
\usepackage{times}
\usepackage{soul}
\usepackage{afterpage}

\setcitestyle{super}

\newcounter{lastnote}

\begin{document}

\sethlcolor{white}

\title{Adsorbate-induced formation of a surface-polarity-driven nonperiodic superstructure}

\author{Chi Ming Yim}
\email{c.m.yim@sjtu.edu.cn}
\affiliation{Tsung Dao Lee Institute \& School of Physics and Astronomy, Shanghai Jiao Tong University, Shanghai, 201210, China}
\affiliation{SUPA, School of Physics and Astronomy, University of St Andrews, North Haugh, St Andrews, Fife, KY16 9SS, United Kingdom}

\author{Yu Zheng}
\affiliation{Tsung Dao Lee Institute \& School of Physics and Astronomy, Shanghai Jiao Tong University, Shanghai, 201210, China}

\author{Olivia R. Armitage}
\affiliation{SUPA, School of Physics and Astronomy, University of St Andrews, North Haugh, St Andrews, Fife, KY16 9SS, United Kingdom}

\author{Dibyashree Chakraborti}
\affiliation{SUPA, School of Physics and Astronomy, University of St Andrews, North Haugh, St Andrews, Fife, KY16 9SS, United Kingdom}
\affiliation{Max Planck Institute for Chemical Physics of Solids, N\"othnitzer Stra\ss e 40, 01187 Dresden, Germany}
\author{Craig J. Wells}
\affiliation{SUPA, School of Physics and Astronomy, University of St Andrews, North Haugh, St Andrews, Fife, KY16 9SS, United Kingdom}
\author{Seunghyun Khim}
\affiliation{Max Planck Institute for Chemical Physics of Solids, N\"othnitzer Stra\ss e 40, 01187 Dresden, Germany}
\author{Andrew Mackenzie}
\affiliation{Max Planck Institute for Chemical Physics of Solids, N\"othnitzer Stra\ss e 40, 01187 Dresden, Germany}
\affiliation{SUPA, School of Physics and Astronomy, University of St Andrews, North Haugh, St Andrews, Fife, KY16 9SS, United Kingdom}
\author{Peter Wahl}
\affiliation{SUPA, School of Physics and Astronomy, University of St Andrews, North Haugh, St Andrews, Fife, KY16 9SS, United Kingdom}
\affiliation{Physikalisches Institut, Universität Bonn, Nussallee 12, 53115 Bonn, Germany}

\date{\today}

\begin{abstract} 
The chemical and electronic properties of surfaces and interfaces are important for many technologically relevant processes, be it in information processing, where interfacial electronic properties are crucial for device performance, or in catalytic processes, which depend on the types and densities of active nucleation sites for chemical reactions. Quasi-periodic and nonperiodic crystalline surfaces offer new opportunities because of their inherent inhomogeneity, resulting in localisation and properties vastly different from those of surfaces described by conventional Bravais lattices. Here, we demonstrate the formation of a nonperiodic tiling structure on the surface of the frustrated antiferromagnet \ce{PdCrO2} due to hydrogen adsorption. The tiling structure exhibits no long-range periodicity but comprises few-atom hexagonally packed domains covering large terraces. Measurement of the local density of states by tunnelling spectroscopy reveals adsorption-driven modifications to the quasi-2D electronic structure of the surface layer, showing exciting opportunities arising from electron localisation.
\end{abstract}


\maketitle

\section*{Introduction}

Surfaces and interfaces play a crucial role in modern technology: from information processing, which occurs largely in semiconductor heterostructures, to large-scale chemical processes facilitated by catalysts. The physics and chemistry of surfaces are as rich as their importance for technological processes: broken symmetry at the surface of a single crystal on its own already leads to novel electronic states and often complex structural reconstructions, providing an opportunity to stabilise entirely different properties at the surface compared to the bulk.

The formation of the surface and concomitant breakage of bonds can result in a range of surface modifications, from mere relaxation, where the surface retains the same symmetries in the directions parallel to the surface as the bulk, to reconstructions, which reduce the symmetry but usually are commensurate with the bulk. Here, we report on the discovery of a new type of nonperiodic tiling structure of hydrogen whose formation is driven by surface polarity in the surface layer from a high-quality single crystal.  Due to its lack of periodicity, this tiling structure cannot be described as a commensurate or incommensurate superstructure nor as a quasi-crystalline surface layer. The observed tiling structure self-assembles with structural elements on length scales of a few nanometres, comparable to those of supra-molecular networks\cite{barth_engineering_2005} however without the need of organic molecules. 
So far, quasi-crystalline, quasiperiodic or amorphous surfaces have been reported on the surface of quasi-crystalline materials\cite{franke_quasicrystalline_2002} or in thin layers grown on a single crystal\cite{forster_quasicrystalline_2013,forster_observation_2016,lewandowski_crystalline_2019, lewandowski_determination_2019,buchner_building_2016} or in 2D polymer networks \cite{alexa_short-range_2019,lackinger_stm_2011}. 

Nonperiodic tilings and quasi-crystals have exciting consequences for the electronic structure\cite{widmer_low-temperature_2006,rogalev_fermi_2015} and potentially for the catalytic, chemical, and mechanical properties\cite{thiel_quasicrystal_2008}. Moir\'e superstructures provide comparable opportunities for local variations in reactivity \cite{zhang_moire_2021}. The significantly more inhomogeneous structure of nonperiodic surfaces compared to that of a normal crystal creates opportunities for tailored sites of different catalytic activities or templates for nucleation of molecules in a range of geometries. Potential applications of these nonperiodic surfaces can go as far as nonstick surfaces for frying pans\cite{rivier_non-stick_1993}.

Here, we study the surface-polarity driven formation of a nonperiodic tiling structure of hydrogen atoms on the Pd-terminated surface of delafossite oxide \ce{PdCrO2} (see Fig.~\ref{Fig-intro}, a and b side-view crystal structures for pristine \ce{PdCrO2} before and after hydrogen adsorption). In the bulk, \ce{PdCrO2} is a frustrated magnetic metal\cite{takatsu_critical_2009}, with highly anisotropic electronic properties\cite{ok_quantum_2013,hicks_quantum_2015}. The interplay between the Mott insulating \ce{CrO2} layers and the metallic Pd layers that host itinerant electronic states makes \ce{PdCrO2} an ideal playground for testing theories of correlated electrons in solids with spectroscopic \cite{sobota_electronic_2013,noh_direct_2014,sunko_probing_2020,yim_avoided_2024} and transport properties \cite{arsenijevic_anomalous_2016,daou_unconventional_2017} and provides exciting opportunities to tailor these through thin-film growth\cite{harada_thin-film_2021}.
The surfaces of many delafossites are polar as a result of the charge distribution within the unit cell. In \ce{PdCrO2}, the \ce{Pd} ions carry a nominal charge of $+1$ and the \ce{CrO2} layers of $-1$ (Fig.~\ref{Fig-intro}a)\cite{mackenzie_properties_2017}. This means that the surface will be susceptible to electronic or structural reconstruction to avoid a polar catastrophe\cite{tasker_stability_1979}. Electronic reconstruction is observed for the Pd and CoO$_2$ terminations of PdCoO$_2$\cite{sunko_maximal_2017,mazzola_itinerant_2018,yim_quasiparticle_2021,kong_fully_2021} and for the CrO$_2$ termination of \ce{PdCrO2}\cite{yim_avoided_2024}, although structural rearrangements are also found\cite{noguera_polar_2000}. Beyond electronic correlation effects, the Pd surface of \ce{PdCoO2} has been reported to exhibit excellent electrocatalytic activity for the hydrogen evolution reaction\cite{li_situ_2019,podjaski_rational_2020}, surpassing nanostructured palladium\cite{moumaneix_interactions_2023,metzroth_accelerating_2021,darmadi_high-performance_2020} and has been found highly reactive, showing evidence of adsorption of gas even in ultra-high vacuum\cite{siemann_dichotomy_2025}. 

\section*{Results}
\subsection*{STM appearance of the Pd-terminated surface of delafossite \ce{PdCrO2}}
We cleaved high-quality single-crystal samples of delafossite \ce{PdCrO2} normal to their \textit{c}-axis at a \textit{nominal} temperature of $\sim 20~\mathrm K$. Due to the crystal structure and the nature of chemical bonding, cleavage occurs typically between the \ce{Pd} and the \ce{CrO2} layers, resulting in two predominant surface terminations: a \ce{Pd}- and a \ce{CrO2}-terminated surface (details on distinguishing between the two surface terminations in scanning tunnelling microscopy (STM) are provided as \hl{Supplementary Notes 1 and 2 and Supplementary Figures~1 and 2}). Here, we concentrate on the \ce{Pd} termination. Figure~\ref{Fig-intro}c shows a large-scale topographic image of a Pd-terminated surface, showing that the surface does not appear flat as seen in PdCoO$_2$ \cite{mazzola_tuneable_2022}, but is characterised by arrays of hexagonal islands of different sizes, shapes, and orientations tightly packed with each other.  As a visual guide, in Fig.~\ref{Fig-intro}d we show the same image overlaid with hexagons of different colours representing islands of different types. The Fourier transformation (Fig.~\ref{Fig-intro}e) shows clear and sharp peaks due to the atomic lattice, less sharp peaks (indicated by a red arrow) due to the spatial arrangement of the hexagonal islands, and other broad features originating from the internal atomic arrangements within those islands.  The lack of a single motif in this nanostructured phase results in a not so well-defined periodicity in the arrangement of the islands, in turn leading to broadening of the related features in the Fourier transformation.  All of these are testament of the nonperiodic nature of the observed tiling structure. The topographic image reveals islands of distinct sizes that we shall henceforth call clusters. 

From a detailed comparison with density functional theory (DFT) calculations (see the Methods section), we find that the observed nonperiodic tiling structure can be understood as a result of dissociative adsorption of hydrogen from the residual vacuum, in which each H atom bonds directly on top of one Pd atom, forming a $(1\times 1)$-H ordered structure within each cluster (see Supplementary Note 3 and Supplementary Figures~3 to 5 for the calculation results in full, and Supplementary Figure~6 for an experimental STM image confirming our assignment of site occupation of hydrogen within the clusters.). Their adsorption behaviour on the Pd-terminated surface of \ce{PdCrO2} observed here is very different from that on the $(111)$ surface of Pd single crystals, where H preferably adsorbs at the hollow sites and their adsorption does not lead to any tiling phase as the one observed here \cite{fernandez-torres_dynamics_2006}. To better illustrate the hydrogen-induced tiling phase observed in STM, we show in Fig.~\ref{Fig-intro}f a schematic model of the tiling structure. The surface exhibits a complex order that consists of characteristic groups of atomic hydrogen that form clusters of different sizes and shapes (as represented by the overlaid hexagons of different colours in Fig.~\ref{Fig-intro}d), and line segments without hydrogen that form the boundary between clusters. These patterns are tiled in a space filling manner in two dimensions, but without resulting in any distinct periodicity, as can be seen in the Fourier transformation of the
topographic image in Fig.~\ref{Fig-intro}e.

We also note that in addition to the $(1\times1)$-H clusters, the surface is also populated with a number of defects that are present in the form of lump-and-hole pairs.  Formed as a result of surface cleavage, these defects account for a surface defect concentration of $\sim 5.3\%$, and therefore have only a minor effect on the overall surface structure (see Supplementary Note 4 and Supplementary Figure~7). 

\subsection*{Spectroscopic signature of hydrogen and the properties of the clusters}
The hydrogen chemisorbed on the Pd surface termination can be detected through its vibrational modes in inelastic tunneling spectroscopy. In Fig.~\ref{Fig-strain}a we present a $\mathrm{d}^2I/\mathrm{d}V^2$ tunneling spectrum taken at the centre of one of the clusters, showing sharp features symmetric around zero bias at energies $|E|=42~\mathrm{meV}$, $84~\mathrm{meV}$ and $272~\mathrm{meV}$, respectively. These are clear signatures of inelastic tunneling. From DFT calculations (see the "Methods" section) we obtain vibrational modes of the chemisorbed hydrogen at energies of $61~\mathrm{meV}$ for the in-plane vibrational mode and $257~\mathrm{meV}$ for the out-of-plane mode, in good agreement with the experiment. Similar values were obtained in previous DFT studies of the surface phonon modes of H/Pt(111) \cite{hong_first-principles_2005}, where the in-plane vibrational mode was found at $47.4~\mathrm{meV}$, and the out-of-plane mode at $277.2~\mathrm{meV}$. 

For the $(1\times1)$-H tiling structure formed on the Pd-terminated surface, it would be expected that the apparent barrier height, a proxy for the work function of the sample, exhibits significant variation between Pd atoms with and without adsorbed hydrogen. We show in Fig.~\ref{Fig-strain}, b and c a topographic $z(\mathbf{r})$ image and a map of the local barrier height $\phi(\mathbf{r})$ taken simultaneously from a small region of the nonperiodic tiling. Also shown is a line cut through the $z(\mathbf{r})$ image and the $\phi(\mathbf{r})$ map across several clusters, see Fig.~\ref{Fig-strain}d. The $\phi(\mathbf{r})$ value varies substantially throughout the tiling, and reaches its maxima $(\sim7~\mathrm{eV})$ and minima $(\sim5~\mathrm{eV})$ at the centre of the clusters and the boundary region, respectively. Consistent with the calculation results \hl{(see Supplementary Note 3 and Supplementary Figure~3 for details)}, our $\phi(\mathbf{r})$ data further support the scenario of electron transfer from the Pd surface to the adsorbed hydrogen, followed by an increase in the surface work function. 

The charge transfer also results in a Coulomb repulsion between the hydrogen atoms and consequently their displacements towards the boundary. From a high-resolution image, we have analysed the lateral positions of the hydrogen atoms within each of the clusters. Fig.~\ref{Fig-strain}e shows a close-up image of the nonperiodic tiling with crosses marking the positions of the \ce{Pd} atoms within the \ce{Pd} surface layer, with their positions determined by numerical fitting using a 2D Gaussian function to every spot on the Pd lattice appearing in the box-filtered image, which was itself generated using only the lattice Bragg peaks in the Fourier transformation, see Supplementary Figure~8 for details. It can be seen that several of the hydrogen atoms in the clusters relax away from their ideal positions (see Fig.~\ref{Fig-strain}, f and g).  Through a detailed analysis of the atomic positions, we find that hydrogen atoms at the periphery of the clusters displace more from their ideal positions than those at the centre (Fig.~\ref{Fig-strain}h). 

We also find that clusters of different sizes and shapes are spatially arranged to have an edge-to-edge separation distance between adjacent clusters of $\sim \sqrt{3}a$, where $a$ is the lattice constant of the \ce{Pd} surface layer. Having such a separation distance also means that the boundary region separating the clusters is one atom wide, confirming its composition of lines of Pd atoms with no adsorbed hydrogen, in turn explaining the less visible appearance of the boundary region in STM. 


\subsection*{Neural Network Characterisation and bias dependent appearance of the nonperiodic tiling phase}

To analyse the spatial arrangement of the $(1\times1)$-H clusters in the surface layer in more detail, we used a neural network to categorise the clusters seen in the STM images according to their types. We trained the neural network to recognise the different arrangements of atoms as shown in Fig.~\ref{Fig-02}a using YOLOv3 (see the Methods section for technical details) from a training data set consisting of hundreds of images of individual clusters. We designated the different types of clusters by a symbol $\mathrm{T}_k$, where $k$ represents the number of hydrogen atoms in the central part of the cluster (also see Supplementary Figure~9 the STM images showing the presence of an adsorbed H atom at the central position of clusters $\mathrm{T}_1$ and $\mathrm{T}_7$). 

Our results (Fig.~\ref{Fig-02}b) from the analysis of a $(50~\mathrm{nm})^2$ image (Supplementary Figure~10) indicate that among the detected clusters, the $\mathrm{T}_2$ clusters are the most common, followed by other cluster types, including $\mathrm{T}_4$, $\mathrm{T}_1$, $\mathrm{T}_3$, and $\mathrm{T}_5$.  The rarest are $\mathrm{T}_7$ clusters. Note that the non-observation of any $\mathrm{T}_6$ cluster is probably due to its lack of symmetry (rotational or mirror) as compared to other cluster types. In particular, the distribution indicates the prevalence of two-fold rotationally ($C_2$) symmetric cluster types $\mathrm{T}_2$ and $\mathrm{T}_4$ with an even number of atoms in the centre.  Taking into account the boundary positions shared among adjacent clusters, we calculated the hydrogen coverage [in monolayer (ML)] within each cluster type [$\theta_{H}(\mathrm{T}_k)$, Fig.~\ref{Fig-02}b], and determined the overall coverage of hydrogen in the imaged area to be $\approx 0.63~\mathrm{ML}$, seemingly slightly overcompensating the surface polarity.  From the results of our analysis, we propose that the number distribution of the clusters is governed by the intricate balance between the hydrogen coverage, repulsive interactions between the adsorbed hydrogen and the possible charge transfer between the adsorbed hydrogen and the substrate (the Pd surface layer). 

The determination of the cluster positions also allows the extraction of their mean distance (Fig.~\ref{Fig-02}c). A Gaussian fit to the histogram reveals a mean distance of $\mu\sim 11.9$\AA~with a standard deviation of $\sigma\sim 0.9$\AA.  Using $\mu$ as the separation distance of the clusters arranged in a perfect hexagonal array, we have calculated the associated wavevector to be $|\mathbf{q}| = 0.61 \pm 0.05$\AA$^{-1}$.  This calculated value agrees perfectly with that arising from the spatial arrangement of the clusters as shown in the Fourier transformation (Fig.~\ref{Fig-intro}e). 

Assuming that all clusters have the same size and shape, and using $\mu$ as the lattice constant of the cluster unit cell, we, by comparison of the areas between the cluster and the primitive unit cell, determined that each cluster on average contains $\sim 16.8$ atoms.  This is very close to the number of atoms $(=16)$ contained within each $\mathrm{T_2}$ cluster, in which atoms in the boundary regions that are shared with adjacent clusters are also counted. 

The formation of the $(1\times1)$-H tiling phase also leads to a strong spatial variation of the local electronic structure acrossing the tiling, as reflected in the drastic differences in the differential conductance spectra $g(V)$ acquired in the centre of a cluster and at a boundary position between the clusters (Fig.~\ref{Fig-02}d), together with the bias dependence of their STM appearance (Fig.~\ref{Fig-02}e), revealing dramatic changes in their apparent height.  Furthermore, comparing the $g(V)$ spectra obtained from the nonperiodic tiling with that of the pristine surface (gray curve), we observe that the sharp conductance peak at $\sim 80~\mathrm{meV}$ found on the pristine surface has shifted to $\sim 310~\mathrm{meV}$ after hydrogen adsorption, which we attribute to electron transfer from the surface to hydrogen, as is also seen from the projected density of states from the DFT calculations \hl{(Supplementary Figure~5)}. 

\subsection*{Electronic confinement in the hydrogen clusters}
 The nonperiodic nature of the observed tiling and the different cluster sizes create an electronic landscape governed by localisation with spatially varying bound state energies due to quantum confinement. The strong influence of the nonperiodic tiling on the electronic structure is readily visible in topographic images acquired at different bias voltages (Fig.~\ref{Fig-02}e). The local confinement occurs on a smaller energy scale and can be visualised by spectroscopic imaging. In Fig.~\ref{Fig-03} we show a topographic image (Fig.~\ref{Fig-03}a) and spectroscopic $g(V,\mathbf{r})$ map slices (Fig.~\ref{Fig-03}, b to d) recorded from a $(5~\mathrm{nm})^2$ region of the $(1\times1)$-H tiling structure, showing that the bound states associated with each of the different types of clusters exhibit an increase in the density of states at different energies. 

To demonstrate this, in Fig.~\ref{Fig-03}, e to h, we show the characteristic tunnelling spectra extracted from cluster types $\mathrm{T_1}$ to $\mathrm{T_4}$.  As shown in Fig.~\ref{Fig-03}e, the spectrum extracted at the corner positions of a $\mathrm{T}_1$ cluster (light blue) exhibits a sharp peak at $E_p=-20~\mathrm{meV}$, which is not present in the spectrum extracted at the central position (dark blue).  This difference is also manifested in the $g(V,\mathbf{r})$ map slice at the peak energy $eV=E_\mathrm p$ (inset of Fig.~\ref{Fig-03}e), where strong maxima are seen at the corners of the $\mathrm{T}_1$ cluster, leading to a shape like a flower with six petals.  Similar resonant states are observed in other cluster types including $\mathrm{T}_2$ and $\mathrm{T}_4$.   In $\mathrm{T}_2$ clusters, the electronic states at $E=-62.5~\mathrm{meV}$ (Fig.~\ref{Fig-03}f) are localised at the cluster corners near the long edge (Fig.~\ref{Fig-03}f).  In $\mathrm{T}_4$ clusters, the states at $-10~\mathrm{meV}$ (Fig.~\ref{Fig-03}h) are localised at the short edges of the cluster, and at the positions near the center of the cluster (inset of Fig.~\ref{Fig-03}h).  However, we do not observe clear localised states within $\mathrm{T}_3$ clusters (see Fig.~\ref{Fig-03}g and its inset), which we speculate to be related to the fact that unlike $\mathrm{T}_{1}$, $\mathrm{T}_{2}$ and $\mathrm{T}_{4}$ clusters that have at least one pair of parallel, equal length edges within each cluster, $\mathrm{T}_{3}$ clusters do not have any such edge pair. Without such an edge pair, strong electronic resonance cannot occur within the cluster. Our results therefore show clear signatures of quantum confinement within the clusters. The fact that the bound states are seen only for negative bias voltages suggests that they originate from a hole-like band that has a band top close to the Fermi energy. 

\section*{Discussion}
Our results show the adsorption induced formation of a nonperiodic tiling structure consisting of $(1\times1)$-H clusters of different types on a perfectly ordered high-purity single-crystal surface. The surfaces of \ce{PdCrO2} are polar surfaces - in the bulk adjacent layers of Pd and \ce{CrO2} exhibit a nominal valence of $+1$ and $-1$, which leaves the bulk charge neutral, but creates a polar catastrophe at the surface unless lifted by a reconstruction, be it electronic or structural. Previous work shows that the Pd surface terminations of both \ce{PdCoO2} and \ce{PdCrO2} exhibit ferromagnetic surface states of highly itinerant electrons \cite{mazzola_itinerant_2018}.  Reaction of the surface with molecular hydrogen from residual vacuum results in a nonperiodic tiling superstructure of chemisorbed hydrogen with properties drastically different from the chemically homogeneous Pd layer. Evidence for hydrogen adsorption has previously been observed for the Pd-termination of \ce{PdCoO2}\cite{siemann_dichotomy_2025}.
This interpretation is supported by measurements of the local density of states and the local barrier height, where we observe large variations between Pd sites with and without adsorbed hydrogen. Our interpretation is also consistent with the significant relaxation of hydrogen within the clusters and the vibrational modes detected within them.  Scans performed at high bias voltage can lead to desorption of hydrogen from the tiling structure and concomitant recovery of a pristine surface \hl{(see Supplementary Figure~11)}. As a result, other possibilities for the formation of the tiling structure, such as the formation of Pd vacancies at the boundary (which would result in significant Pd deposits elsewhere on the surface), are naturally ruled out.

The formation of the tiling structure of hydrogen in the \ce{Pd} surface layer of \ce{PdCrO2} also results in a localisation of the electronic states close to the Fermi energy due to quantum confinement, confirmed through our spectroscopic investigations: the electronic structure varies significantly between different types of clusters, showing strong resonant states in the vicinity of the Fermi energy. 

Furthermore, in addition to confirming that the composition of the tiling structure depends on the temperature at which it forms, we also note that the tiling structure is absent from the $12~\mathrm{K}$-cleaved sample, which itself reveals a metallic surface (see Supplementary Figure~12) with similar spectroscopic signatures as seen on the Pd-terminated surface of \ce{PdCoO2}\cite{mazzola_tuneable_2022}.  We attribute the absence of the tiling on the $12~\mathrm{K}$-cleaved sample to the huge reduction of the hydrogen partial pressure inside the vacuum chamber as a result of cold trapping by the liquid helium-cooled surfaces during sample cleavage, in turn leading to no adsorption of hydrogen on the sample surface.

The electronic and structural inhomogeneity of the adsorption-induced tiling structure on the Pd-termination reveals new opportunities to design surfaces and template substrates with a range of catalytically active sites, with targeted reactivity based on geometry and local electronic structure. 
An interesting open question is how hydrogen adsorption affects the strongly correlated electronic states in the Mott insulating \ce{CrO2} layer. The anisotropy of the electronic transport properties of the material, together with the highly two-dimensional nature and extremely high chemical reactivity of the Pd-termination may provide important clues to understand the excellent electrocatalytic activity of some delafossites for the hydrogen evolution reaction \cite{li_situ_2019,podjaski_rational_2020} and underpin future rational design of delafossite-based catalysts that take advantage not only of the electronic configuration but also the nonperiodic nature of the surface structure.



\section*{Methods}
\textbf{Scanning tunneling microscopy/spectroscopy (STM/S) measurements.}  The STM/S measurements were performed using a home-built low temperature STM that operates at a base temperature of $1.8~\mathrm{K}$ \cite{white_stiff_2011}.  \ce{Pt/Ir} tips were used, and were conditioned by field emission performed on a \ce{Au} target before use.  Differential conductance $\mathrm{d}I/\mathrm{d}V$ or $g(V)$ and inelastic tunneling $\mathrm{d}^2I/\mathrm{d}V^2$ spectra, and spectroscopic maps were recorded using a standard lock-in technique, with the frequency of the bias modulation set at $413~\mathrm{Hz}$.   All reported data were obtained at a sample temperature of $4.2~\mathrm{K}$ unless otherwise stated.  For STM/S measurements, the clean sample surfaces were prepared by \textit{in-situ} cleaving at a \textit{nominal} sample temperature of $\sim 20~\mathrm{K}$.

To examine the reproducibility of the nonperiodic tiling structure, similar STM/S measurements were performed using a commercial \textit{Unisoku} USM1300 ultrahigh vacuum (UHV) STM machine that can operate at a base temperature of $300 ~\mathrm{mK}$, where clean sample surfaces were prepared by \textit{in-situ} cleaving of the samples held at sample temperatures of $78~\mathrm{K}$ or $12~\mathrm{K}$ under the UHV condition (base pressure $~7\times 10^{-11} ~\mathrm{mbar}$).  In the experiments, we observed a similar tiling phase formed on the \ce{Pd} terminated surface of the sample cleaved at $78~\mathrm{K}$, but a pristine, unreconstructed \ce{Pd} surface of the sample cleaved at $12~\mathrm{K}$.\\ 

\textbf{Crystal growth.} Single-crystal samples of \ce{PdCrO2} were grown by the \ce{NaCl}-flux method as reported in Ref.~\onlinecite{takatsu_single_2010}. First, polycrystalline \ce{PdCrO2} powder was prepared from the following reaction at $960^\circ\mathrm{C}$ for four days in an evacuated quartz ampoule: 
\begin{equation}
    \ce{2LiCrO2 + Pd + PdCl2 -> 2PdCrO2 + 2LiCl}.
\end{equation} 
The obtained powder was washed with water and aqua regia to remove \ce{LiCl}. The polycrystalline \ce{PdCrO2} and \ce{NaCl} were mixed in the molar ratio of 1:10. Then, the mixture in a sealed quartz tube was heated at $900^\circ\mathrm{C}$ and slowly cooled down to $750^\circ\mathrm{C}$. \ce{PdCrO2} single crystals were harvested after dissolving the \ce{NaCl} flux with water.\\

\textbf{Neural network analysis.}  The YOLOv3 neural network was trained using codes from Ref.~\cite{yolov3}, to identify the six most common cluster types using the images containing individual clusters of each cluster type.  The training and testing data sets consisted of eight and five images respectively for each class.  To ensure that the network was trained on clusters at different angular orientations, to expand the data set, copies of each image were rotated at $60^{\circ}$ intervals. This resulted in a total of $282$ images used for training and $174$ for testing.  The $z$ values in the images were re-scaled to have a zero mean ($\overline{z}$) and unity standard deviation ($\sigma_{z}$).  When applying the neural network to the $(50~\mathrm{nm})^2$ image, rescaled $z$ values that exceed $3\sigma_{z}$ from $\overline{z}$ on both sides were excluded due to the presence of surface defects. The rescaling and exclusion of the extreme values were then repeated to produce the final image. This contrast saturation approach provided better contrast for the surface clusters, in turn allowing for a higher detection rate by the network.

\textbf{DFT calculations.}  To determine the origin of the nanocluster network and valence disproportionation, we have performed extensive DFT calculations to model a number of scenarios. Ultimately, the one that best matches the experiments and captures all observations is that the valence change is induced by hydrogen adsorbed on the Pd surface, resulting in a local change of the work function and significant local modification of the electronic structure of the Pd layer through hole doping.
The calculations have been performed on a two-layer slab of \ce{PdCrO2} with a range of unit cell sizes. Scenarios that we have investigated include a vacancy network, adsorption of hydrogen and carbon monoxide, and structural reconstructions.
For the calculations, we have added a Coulomb repulsion $U$ for the $d$-orbitals of the Cr, consistent with previous studies\cite{yim_avoided_2024}, however note that the inclusion of the $U$ term does not result in any significant differences for the surface layer.
DFT calculations have been performed using PBE functionals using VASP\cite{kresse_ab_1993,kresse_ab_1994,kresse_norm-conserving_1994,kresse_efficiency_1996,kresse_efficient_1996,kresse_ultrasoft_1999} with a plane wave energy cut-off of $800~\mathrm{eV}$. Most calculations have been performed for the unit cell of a $\mathrm{T}_2$ cluster, with a $2\times 2$ k-grid. For structural relaxations, the bottom unit of \ce{PdCrO2} and the bottom Pd surface have been fixed. The bottom surface of the slab has been saturated with the same number of hydrogen atoms as the top surface so that both the bottom and top surfaces have the same chemical configuration.

\section*{Data availability}
The data that support the findings of this study are openly available at https://cstr.cn/32010.11.sjtu.scidata.00000050.\\



\pagebreak
\noindent{\bf Acknowledgements:} We gratefully acknowledge discussions of hydrogen adsorption with Phil King and of preliminary calculations with Chiara Gattinoni, as well as discussions with Federico Mazzola and Gesa Siemann. We further thank Neville Richardson for critical reading of and valuable comments on the manuscript. C.M.Y. and P.W. acknowledge support from EPSRC (EP/S005005/1).  C.M.Y. acknowledges additional support from the Ministry of Science and Technology of China (2022YFA1402702) and TDLI Start-up Fund.  A.P.M. acknowledges support from the MPG. This work used the Cirrus UK National Tier-2 HPC Service at EPCC (http://www.cirrus.ac.uk) funded by the University of Edinburgh and EPSRC (EP/P020267/1), as well as the HPC cluster Hypatia of the University of St Andrews.\\
\noindent{\bf Author contributions:} C.M.Y., P.W. and A.P.M. conceived the project. C.M.Y., Y. Z. and D.C. performed STM experiments. C.M.Y. analysed the data and prepared the figures. O.R.A. performed neural network analysis. C.J.W. prepared the training data set. S.K. and A.P.M. grew the crystals. P.W. performed DFT calculations. C.M.Y., P.W. and A.P.M. wrote the manuscript. All authors discussed and contributed to the manuscript.\\
\noindent{\bf Competing Interests:} The authors declare no competing interests.\\
\noindent{\bf Correspondence:} Correspondence and requests for materials should be addressed to C.M.Y. (email: c.m.yim@sjtu.edu.cn).


\newpage

\begin{figure}
    \caption{\textbf{Hydrogen-adsorption induced formation of a nonperiodic tiling phase on the Pd terminated surface of delafossite \ce{PdCrO2}.}  \textbf{a}, Side-view crystal structure of delafossite oxide \ce{PdCrO2}.  Inside the bulk, \ce{Pd} layers possess electrical charge of $+1$ per Pd atom, layers of \ce{PdCrO2} contain charge of $-1$ per CrO$_2$ octahedra.  Upon cleaving, the vacuum-exposed Pd surface layer possesses charge of $+0.5$ per \ce{Pd} atom.  \textbf{b} As (\textbf{a}), following dissociative adsorption of hydrogen, leading to a nonperiodic tiling phase consisting of $(1\times1)$-H clusters formed at the surface.  \textbf{c}, Topographic STM image of the Pd-terminated surface of \ce{PdCrO2} [$V=-20~\mathrm{mV}$,  $I=500~\mathrm{fA}$; image size: $(20~\mathrm{nm})^2$]. \textbf{d}, As (\textbf{c}), overlaid with an array of hexagons of different colours representing clusters different types distributed on the Pd terminated surface. \textbf{e}, Fourier transformation from a $(50~\mathrm{nm})^2$ area of the nonperiodic tiling phase. Blue circles mark the Bragg peaks associated with the lattice of \ce{Pd} atoms.  A red arrow indicates the wave-vector, $\mathbf{q_{avg}}$, associated with the spatial arrangement of the 2D hexagonal clusters.  The overlaid red dashed circle has radius ($\sim 0.61$\AA $^{-1}$) equal to $|\mathbf{q_{avg}}|$. \textbf{f}, Structural model of the Pd terminated surface layer observed in this study.  The surface is chemi-sorbed by hydrogen, leading to a tiling phase that lacks any periodicity. In the model, green spheres represent atomic H adsorbed on each Pd atoms in the surface layer.}
    \label{Fig-intro}
\end{figure}

\begin{figure}
    \caption{\textbf{Chemical signatures of the $(1\times1)$-\ce{H} clusters} \textbf{a}, Point inelastic tunnelling ($\mathrm{d}^2 I/\mathrm{d} V^2$) spectrum recorded from the central position of a $(1\times1)$-\ce{H} cluster ($V_\mathrm{s}=500~\mathrm{mV}$, $I_\mathrm{s}=100~\mathrm{pA}$; $V_\mathrm{mod}=5~\mathrm{mV}$; 10 sweep average).  The spectrum shows three sets of inelastic peaks at $|E|$ of $42~\mathrm{meV}$, $84~\mathrm{meV}$ (which is the second harmonic of the $42~\mathrm{meV}$ peak) and $272~\mathrm{meV}$ respectively.  Insets: schematics showing the parallel and perpendicular phonon modes of the \ce{H-Pd} bond at $42~\mathrm{meV}$ and $272~\mathrm{meV}$ respectively. \textbf{b}, STM topographic $z(\mathbf{r})$ image of the nonperiodic tiling [$V=200~\mathrm{mV}$,  $I=10~\mathrm{pA}$; image size: $(5~\mathrm{nm})^2$].  \textbf{c}, Corresponding local barrier height $\phi(\mathbf{r})$ map extracted from the $I(\mathbf{r},z)$ spectroscopic imaging data recorded simultaneously with (\textbf{b}).  \textbf{d}, Topographic (light-) and $\phi$ (dark-purple) line-cuts taken along the same line in (\textbf{b}) and (\textbf{c}).  Arrows indicate the positions at the boundary at which $\phi$ reaches local minima.  \textbf{e}, Another $z(\mathbf{r})$ image of the $(1\times1)$-\ce{H} tiling phase [$V=32~\mathrm{mV}$, $I=32~\mathrm{pA}$; image size: $(4~\mathrm{nm})^2$].  \textbf{f}-\textbf{g}, Close-ups of two individual clusters extracted from (\textbf{e}) [image size: $(1.5~\mathrm{nm})^2$].  In (\textbf{e})-(\textbf{g}), crosses mark the positions of \ce{Pd} atoms in the perfect hexagonal lattice.  In (\textbf{f})-(\textbf{g}), red circles and blue squares mark the positions of hydrogen atoms at the center and at the edges of the $(1\times1)$-H clusters respectively. \textbf{h}, Histograms of the in-plane displacement of the hydrogen atoms (relative to their beneath \ce{Pd} atoms) at the center (red) and edges (blue) of the hexagonal clusters in (\textbf{f})-(\textbf{g}), as well as of one other cluster (not shown).  
    }
    \label{Fig-strain}
\end{figure}

\begin{figure}
    \caption{\textbf{Distribution and spatial arrangement of the hexagonal clusters, and their bias-dependent appearance in STM.} \textbf{a}, Zoom-in topographic images of the six most popular $(1\times1)$-\ce{H} cluster types present within the nonperiodic tiling [$V=50~\mathrm{mV}$, $I=1~\mathrm{pA}$; image size: $(2~\mathrm{nm})^2$].  Symbols $\mathrm{T}_k$ denote different cluster types, where $k$ in the subscript denotes the number of hydrogen atoms present in the central part in each cluster type.  The rotation symmetry possessed by each cluster type is also given.  \textbf{b}, Number distribution of the different cluster types, calculated from a $(50~\mathrm{nm})^2$ image (Supplementary Fig.~10) of the nonperiodic tiling using the YOLOv3 neural network approach.  Threshold for detection = $0.7$.  Hydrogen coverage within each cluster type is also plotted, with the overall hydrogen coverage in the $(50~\mathrm{nm})^2$ image calculated to be $\approx0.63~\mathrm{ML}$.  \textbf{c}, Histogram of the average separation distance between neighbouring clusters, extracted from their predicted locations within the same image. A Gaussian fit to the histogram yields mean and standard deviation values of $\mu = 11.87\pm 0.01$\AA~and $\sigma = 0.93\pm 0.01$\AA~respectively. \textbf{d}, Point $g(V)$ spectra recorded at the centre of one $(1\times1)$-\ce{H} cluster and at a position within the boundary region ($V_\mathrm{s}=-1.25~\mathrm{V}$, $I_\mathrm{s}=1~\mathrm{pA}$; $V_\mathrm{mod}=10~\mathrm{mV}$). For comparison, the spectrum recorded from a defect-free position on the pristine Pd-terminated surface of \ce{PdCrO2} ($V_\mathrm{s}=1~\mathrm{V}$, $I_\mathrm{s}=500~\mathrm{pA}$; $V_\mathrm{mod}=10~\mathrm{mV}$) is also provided and is vertically offset for clarity.  A red arrow indicates a shift in energy position of the sharp $\mathrm{d}I/\mathrm{d}V$ at $80~\mathrm{meV}$ to $310~\mathrm{meV}$ before and after adsorption of hydrogen.   Inset of (\textbf{d}), $(8~\mathrm{nm}^2)$ STM topographic image taken from the pristine Pd-terminated surface of \ce{PdCrO2} ($V=200~\mathrm{mV}$, $I=50~\mathrm{pA}$; scale bar: $2~\mathrm{nm}$).  The cross indicates the position at which the grey spectrum in (\textbf{d}) were taken. \textbf{e}, Bias-dependent topographic images taken at the same position on the nonperiodic tiling [image size: $(4~\mathrm{nm})^2$].  The bias voltages at which the images were recorded are indicated by black arrows in (\textbf{d}).
    }
    \label{Fig-02}
\end{figure}

\begin{figure}
    \caption{\textbf{Spatial confinement induced electronic bound states formed within the $(1\times1)$-\ce{H} clusters.}  \textbf{a},  STM topographic image of the nonperiodic tiling structure [$V=-200~\mathrm{mV}$, $I=600~\mathrm{fA}$; image size: $(5~\mathrm{nm})^2$].  \textbf{b}-\textbf{d}, $g(\mathbf{r},V)$ map slices at different energies simultaneously recorded with (\textbf{a}): (\textbf{b}) $-62~\mathrm{meV}$, (\textbf{c}) $-20~\mathrm{meV}$, and (\textbf{d}) $-10~\mathrm{meV}$ ($V_\mathrm{s}=-200~\mathrm{mV}$, $I_\mathrm{s}=50~\mathrm{pA}$;
    $V_\mathrm{mod}=2.5~\mathrm{mV}$).  In (\textbf{a})-(\textbf{d}), circles of different colours mark the $(1\times1)$-\ce{H} clusters of different shapes: $\mathrm{T}_1$ to $\mathrm{T}_4$. \textbf{e}-\textbf{h}, Point $g(V)$ spectra extracted from different positions of each of the clusters $\mathrm{T}_1$ to $\mathrm{T}_4$.  Insets of (\textbf{e})-(\textbf{h}), Topographic images of the clusters [image size: $(1.5~\mathrm{nm})^2$], overlaid with markers indicating where the positions at which the $g(V)$ spectra were extracted.
    }
    \label{Fig-03}
\end{figure}

\clearpage
\pagebreak
\newpage
\begin{center}
\includegraphics[width=\columnwidth]{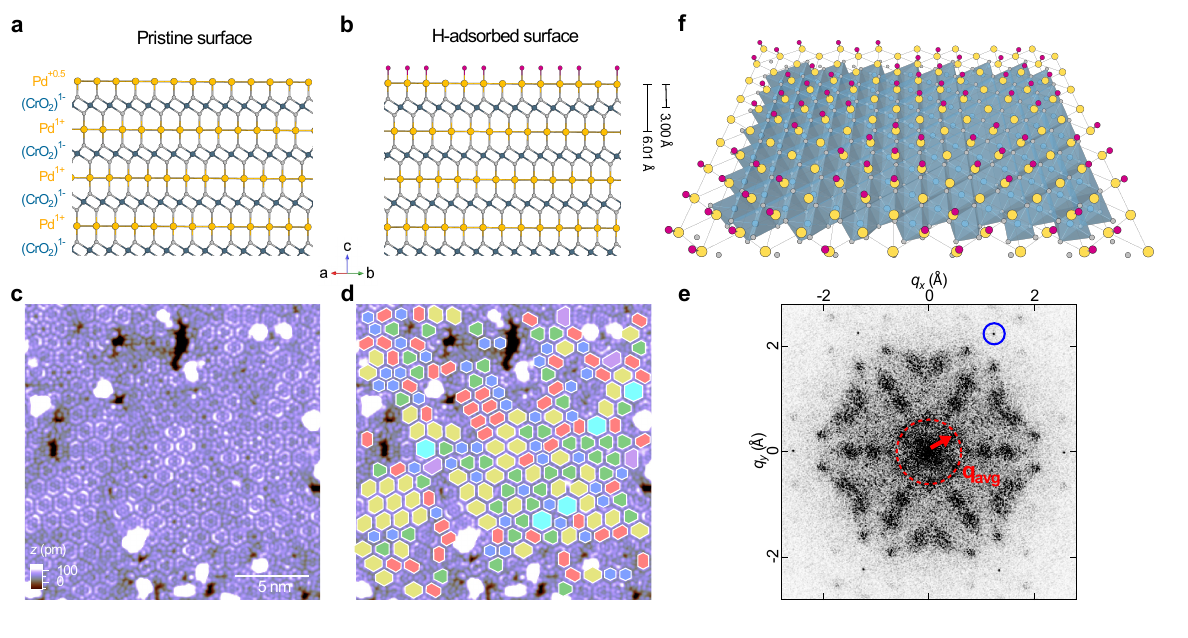}
\end{center}
\huge{Fig. 1}
\newpage
\begin{center}
\includegraphics[width=\columnwidth]{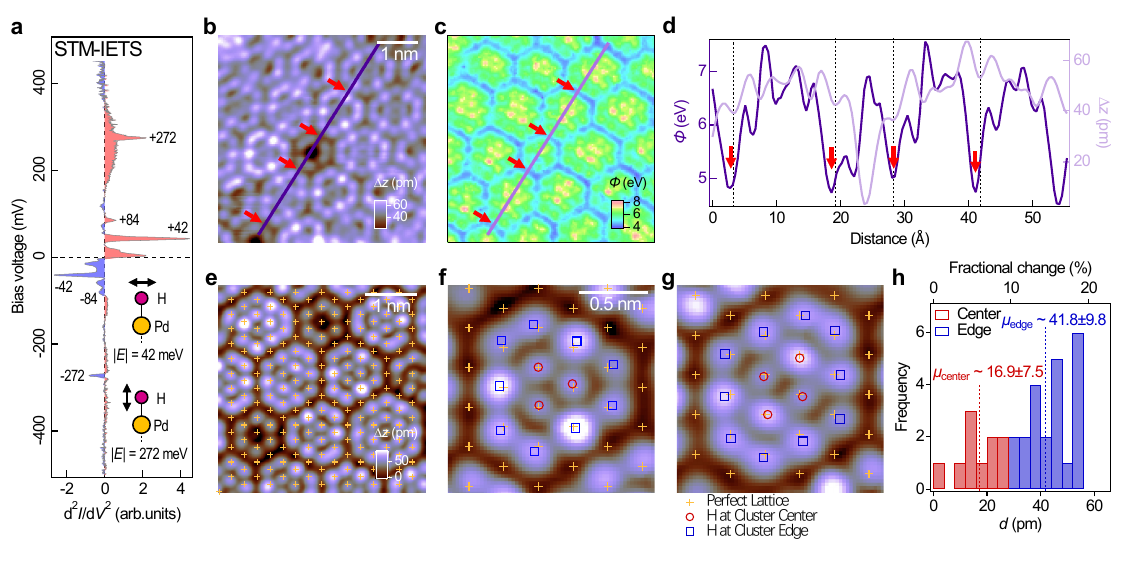}
\end{center}
\huge{Fig. 2}
\newpage
\begin{center}
\includegraphics[width=\columnwidth]{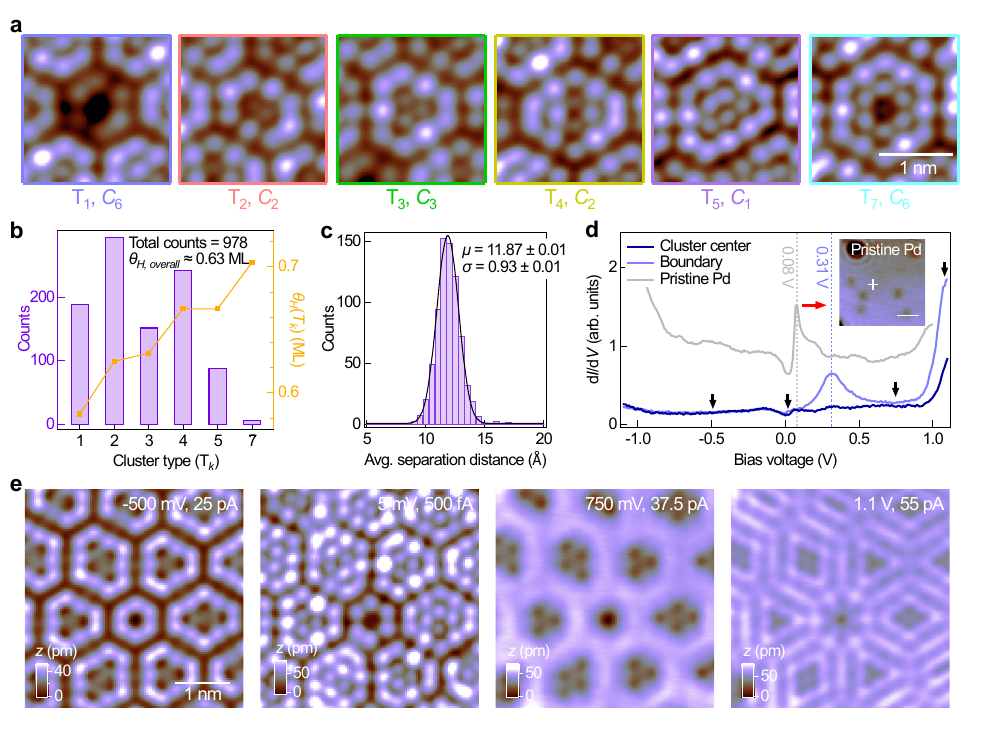}
\end{center}
\huge{Fig. 3}
\newpage
\begin{center}
\includegraphics[width=\columnwidth]{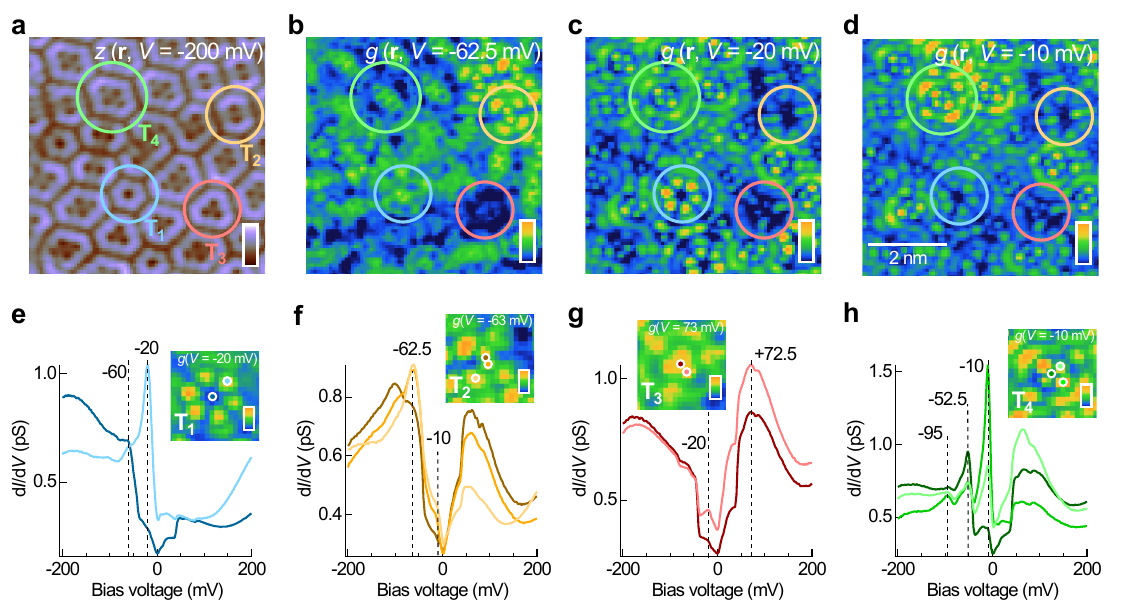}
\end{center}
\huge{Fig. 4}

\end{document}